\documentclass[aps,prl,showpacs,twocolumn,amsmath,amssymb,superscriptaddress]{revtex4-1}
\usepackage{graphicx}
\usepackage{dcolumn}
\usepackage{bm}
\usepackage{units}
\usepackage{upgreek}
\usepackage{ucs}
\usepackage{color}
\usepackage{cancel}
\usepackage{hyperref}
\usepackage[all]{hypcap}
\usepackage{braket}

\begin{document}

\title{Non-equilibrium dynamics in dissipative Bose-Hubbard chains}
\author{Georgios Kordas}
\affiliation{University of Athens, Physics Department, Nuclear \& Particle Physics Section, Panepistimiopolis, Ilissia 15771, Athens, Greece}
\author{Dirk Witthaut}
\affiliation{Forschungszentrum J\"ulich, Institute for Energy and Climate Research (IEK-STE), 52428 J\"ulich, Germany}
\affiliation{nstitute for Theoretical Physics, University of Cologne, 50937 K\"oln, Germany}
\author{Sandro Wimberger}
\affiliation{Dipartimento di Fisica e Science della Terra, Universit\`{a} di Parma Via G.P. Usberti 7/a, 43124 Parma, Italy}
\affiliation{INFN, Sezione di Milano Bicocca, Gruppo Collegato di Parma, Italy}
\email{sandromarcel.wimberger@unipr.it}

\begin{abstract}
Open many-body quantum systems have recently gained renewed interest in the context of quantum information science and quantum transport with biological clusters and ultracold atomic gases. We present a series of results in diverse setups based on a Master equation approach to describe the dissipative dynamics of ultracold bosons in a one-dimensional lattice. We predict the creation of mesoscopic stable many-body structures in the lattice and study the non-equilibrium transport of neutral atoms in the regime of strong and weak interactions.
\end{abstract}

\maketitle
\section{Introduction}\label{sec:01}

Only recently it has been realized that dissipation can be used 
to steer the dynamics of complex quantum systems if it can be 
accurately controlled. Controlled dissipative processes allow 
one, for instance, to prepare pure states for quantum computation \cite{Zoller2008} or 
to implement universal quantum computation \cite{Cirac2009} and to generate deterministically highly entangled states 
\cite{Nature_new, kordas12}. Moreover, dissipation is conjectured to be responsible of coherently enhanced
transport observed in biological clusters, see e.g. \cite{Plenio2013}.

In this paper we investigate the impact of opening a many-body quantum 
system on its dynamical evolution. We will see how dissipation together with
strong interparticle interaction can be used to actively create stable and coherent 
many-body structures. Furthermore, we study simple models for the 
non-equilibrium transport of interacting bosons across quantum-dot like potentials. Driven
by the experimental advance in the implementation of such systems with ultracold atoms 
\cite{Brantut2012, Eckel2014, Krinner2014, Ott_new}, we show how noise and coupling to lead-like 
channels can lead to complex particle transport in one-dimensional chains of quantum wells.

\section{Dissipative and noisy Bose-Hubbard models}\label{sec:02}

The dynamics of ultracold bosonic atoms in a deep optical lattice is given by the well-known Bose-Hubbard (BH) Hamiltonian
\cite{BDZ2008}
\begin{eqnarray}\label{eq:BH1}
\nonumber\hat{H}_{\rm BH} &=& \sum_j \varepsilon_j~\hat{\alpha}^\dag_j \hat{\alpha}_j -
J\sum_j (\hat{\alpha}^\dag_{j+1} \hat{\alpha}_j + \hat{\alpha}^\dag_{j} \hat{\alpha}_{j+1})\\
&&+ \frac{U}{2}\sum_j\hat{\alpha}^\dag_j \hat{\alpha}^\dag_j \hat{\alpha}_j\hat{\alpha}_j.
\end{eqnarray}
$\hat{\alpha}_j$ and $\hat{\alpha}_j^\dagger$ are the bosonic annihilation and
creation operators in mode $j$, $U$ denotes the interaction strength and $J$ the tunneling
strength between the wells. We set $\hbar=1$ throughout, measuring all energies in frequency units.

In the presence of dissipation the dynamics is usually given by a master equation in Lindblad form~\cite{Breu07,Anglin1997,kepe12,witt08, witt09,witt11,trim11,kordas12,kordas13,kollath2011,Kordas2015}
\begin{equation}\label{eq:DBH}
 \frac{d}{dt}\hat{\rho} = -i[\hat{H}_{\rm BH},\hat{\rho}] + \mathcal{L}\hat{\rho},
\end{equation}
where the last term, known as Liouvillian, describes the non-unitary part of the dynamics.

\begin{figure}
  \includegraphics[width=\columnwidth]{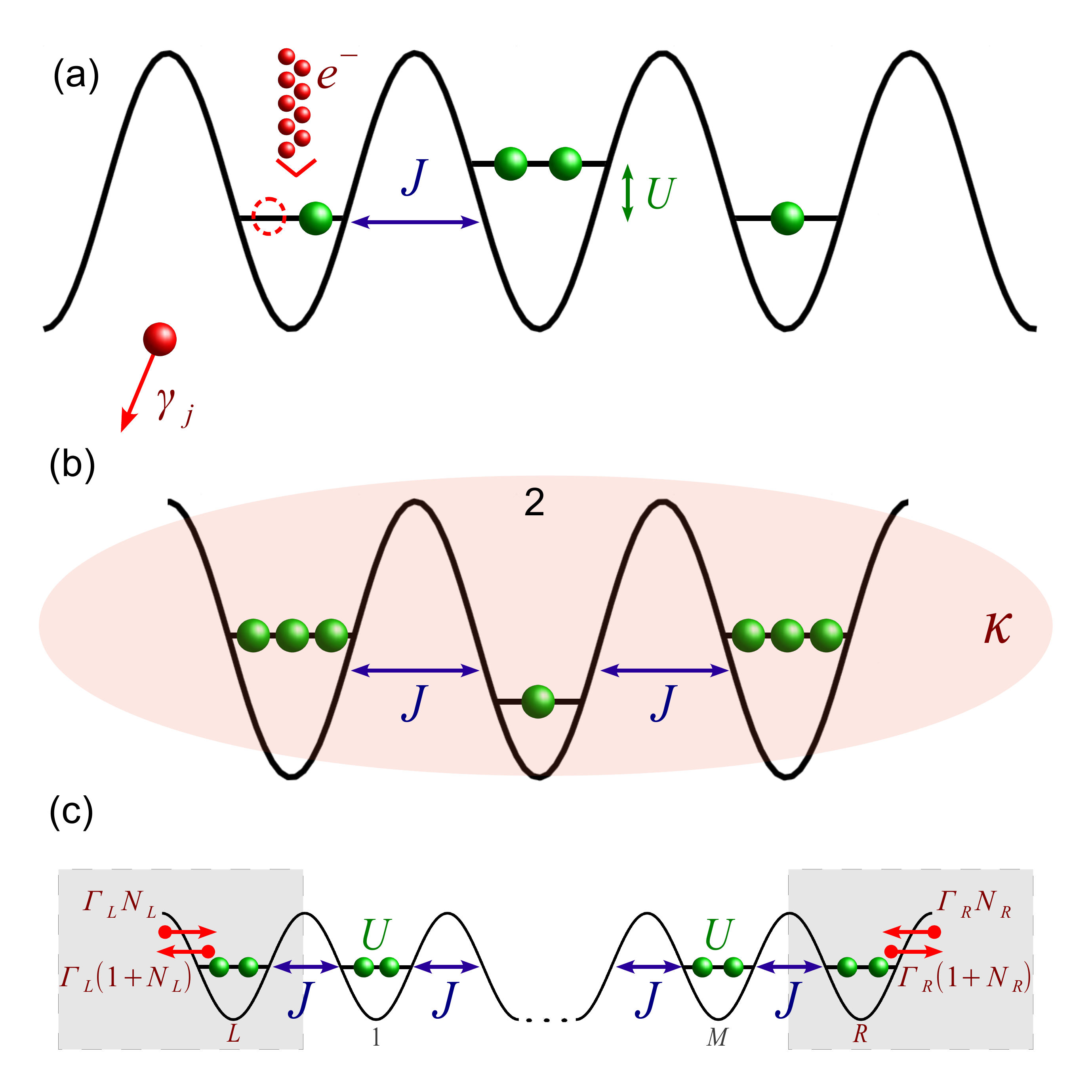}%
  \caption{\label{fig:01}
    Sketches of (a) a Bose-Hubbard chain with localized single particle loss and of two non-equilibrium scenarios: (b) the refilling dynamics of lattice site 2 in the presence of noise and (c) the transport across a chain of quantum dots.}
\end{figure}

In the present paper we will use three different model systems.
In the first one, we will study the dynamics in an optical lattice which is subject to localized
single particle loss, see Fig. \ref{fig:01} (a). Such loss channels are realized in state-of-the-art experiments
\cite{Ott08, Ott09, Laser, Ott2013}. The Liouvillian for this type of dissipation has the form
\begin{equation}\label{eq:L-loss}
 \mathcal{L}_{\rm loss}\hat{\rho} = - \sum_j \frac{\gamma_j}{2}(\hat{\alpha}^\dag_j \hat{\alpha}_j\hat{\rho} +
 \hat{\rho} \hat{\alpha}^\dag_j \hat{\alpha}_j - 2 \hat{\alpha}_j  \hat{\rho} \hat{\alpha}_j^\dag),
\end{equation}
where $\gamma_j$ denotes the loss rate at site $j$. We will also study the effect of global phase noise 
on the non-equilibrium dynamics, which is typically modelled by the following Liouvillian
\begin{equation}\label{eq:L-noise}
 \mathcal{L}_{\rm noise}\hat{\rho} = - \frac{\kappa}{2} \sum_j (\hat{n}^2_j\hat{\rho} +
 \hat{\rho} \hat{n}^2_j - 2 \hat{n}_j  \hat{\rho} \hat{n}_j).
\end{equation}
Here, $\hat{n}_j$ is the local number operator and $\kappa$ determines the strength of the noise, 
which in the experiment may arise from scattering with atoms from the thermal cloud or other density 
dependent random processes \cite{Anglin1997, Zoller2010}. The impact of dephasing on the tunneling dynamics
into an empty lattice site, see Fig. \ref{fig:01} (b), is studied in Sec.~\ref{sec:05}.

In the third setup, we use appropriate creation and destruction processes at the two ends of a BH chain in order to create
a steady state current through the chain, see Fig. \ref{fig:01} (c). We will discuss in detail this system
in Sec.~\ref{sec:06}. In this non-equilibrium scenario
the Liouvillian is given by
\begin{eqnarray}
\nonumber\mathcal{L}_{\rm tr}\hat{\rho} &=& -\sum_{j=R,L}\Bigg\{ \frac{\Gamma_j(1+N_j)}{2}(\hat{\alpha}_j^\dagger \hat{\alpha}_j \hat{\rho} + \hat{\rho}\hat{\alpha}_j^\dagger \hat{\alpha}_j - 2 \hat{\alpha}_j\hat{\rho}\hat{\alpha}_j^\dagger)\\
&&+\frac{\Gamma_j N_j}{2}(\hat{\alpha}_j \hat{\alpha}_j^\dagger \hat{\rho} + \hat{\rho}\hat{\alpha}_j \hat{\alpha}_j^\dagger - 2 \hat{\alpha}_j^\dagger\hat{\rho}\hat{\alpha}_j)\Bigg\},
\end{eqnarray}
where the first term in the right hand side destroys particles with rate $\Gamma_j(1+N_j)$, while the second creates particles with rate $\Gamma_j N_j$, in the left ($j=L$) and right ($j=R$) end of the lattice, see Sec. \ref{sec:06} for further details. Related transport scenarios, just so far without a lattice structure, are being implemented with neutral atoms at the moment \cite{Brantut2012, Eckel2014, Krinner2014}, see also \cite{Nat_Phys, Gattobigio2011, Aspect2006} for experimental injection techniques and  \cite{Ott_new, Cali2012} for preliminary experiments with lattice traps.

\section{Discrete breather formation}\label{sec:03}

In~\cite{witt11,trim11} it was shown that localized single particle losses can be used to create stable nonlinear structures.
For instance, a discrete breather can emerge in a lattice with boundary losses or a coherent dark soliton can be engineered with the help
of phase imprinting and localized losses. Here we will create a discrete breather in the lattice site we desire by using
moving loss. Discrete breathers are spatially localized, time periodic, stable excitations in a perfectly periodic discrete system.
Their existence is a result of the discreteness and nonlinearity of the system \cite{Camp04,Flac08}.

\begin{figure}
  \includegraphics[width=\columnwidth]{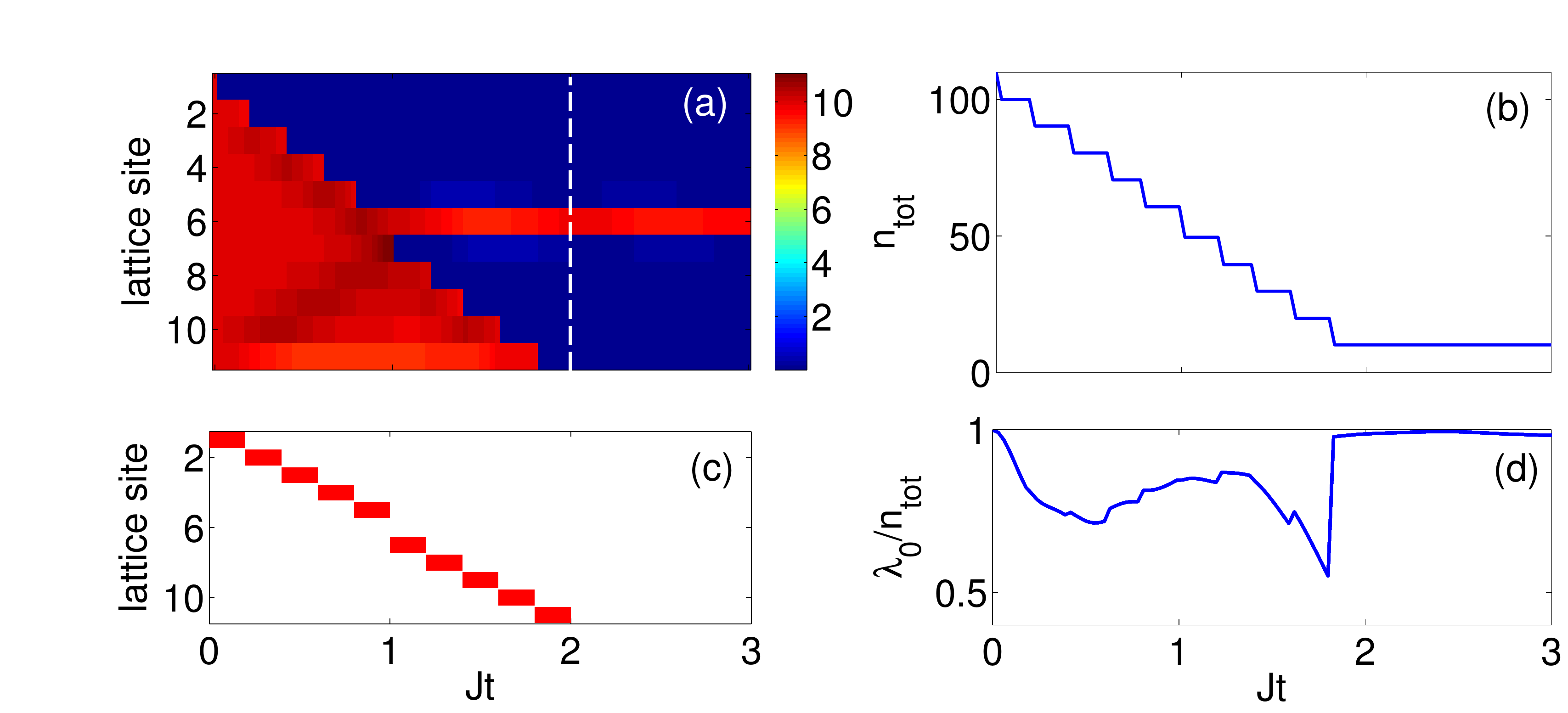}%
  \caption{\label{fig:02}
    Emergence of a discrete breather in a lattice with 11 sites which is subject to moving losses.
 (a) Evolution of the atom density for a large loss rate $\gamma=1200J$. (b) Evolution of the total particle number. (c) The red line
indicates the position of the leaky site as a function of time, the loss stops at $t=2J$. (d) The evolution of the condensate fraction $\lambda_0/n_{\rm tot}$. The interaction strength is $U=J$ and the initial population $n_{\rm tot}(0)=110$.
For the simulations we used the \emph{BBR} method.}
\end{figure}

Our example system consists of a lattice with $M=11$ sites, which is loaded with a pure homogeneous BEC with initial average density of 10 atoms per site. The particle loss can be implemented, for example, with an electron beam, which can be moved very fast with respect to the system time scales \cite{Ott08, Ott09, Ott2013}. For the following, we suppose that the loss starts on the 1st site, and is then scanned through the lattice, as illustrated in Fig. \ref{fig:02} (c). No loss should occur in the site at which we desire to create the breather. The electron beam should ``jump'' this lattice site. To perform the simulation in this rather large many-body system we  used the Bogoliubov back-reaction (BBR) method. This beyond mean-field technique is reviewed in~\cite{witt11,trim11} for our type of problem. It proved to describe well the many-body dynamics in the presence of strong dissipation and allows us to access coherence measures for the system.

In order to create a breather the interaction strength $U$ and the particle loss rate $\gamma$ should be much larger than the tunneling
strength $J$. The interactions should be large enough to induce self-trapping on the non-leaky site, while the loss
should be larger than $J$ in order to avoid substantial tunneling to the neighboring leaky sites. The whole process is depicted in Fig. \ref{fig:02} (a,b), which shows the evolution of the particle density in each lattice site and the total particle number, respectively.
The particles are rapidly removed from the lattice, except for the central non-leaky well, where we find a stable population of particles. Furthermore, the population remains stable on this site even if we switch of the losses completely (at $t=2J$).
In Fig. \ref{fig:02} (d) we have plotted the evolution of the condensate fraction. Due to the strong interactions we observe depletion
of the BEC, but after $t=2J$ we have repurification: an almost pure BEC is localized in the central site (with approximate filling $n_{\rm tot}(0)/M=10$) and non-coherent oscillations are largely damped out. This localized state is stable on all experimentally relevant time scales due to self-trapping in the one-dimensional lattice.

\section{Dissipation induced entanglement}\label{sec:04}

Particle dissipation obviously reduces the total particle number in the lattice, but, as we discussed in the
previous section and in \cite{witt08, witt09, witt11, trim11, kordas12, kordas13}, the loss -- in cooperation with strong interactions -- triggers the formation of interesting (meta)stable structures. 

\begin{figure}
  \includegraphics[width=\columnwidth]{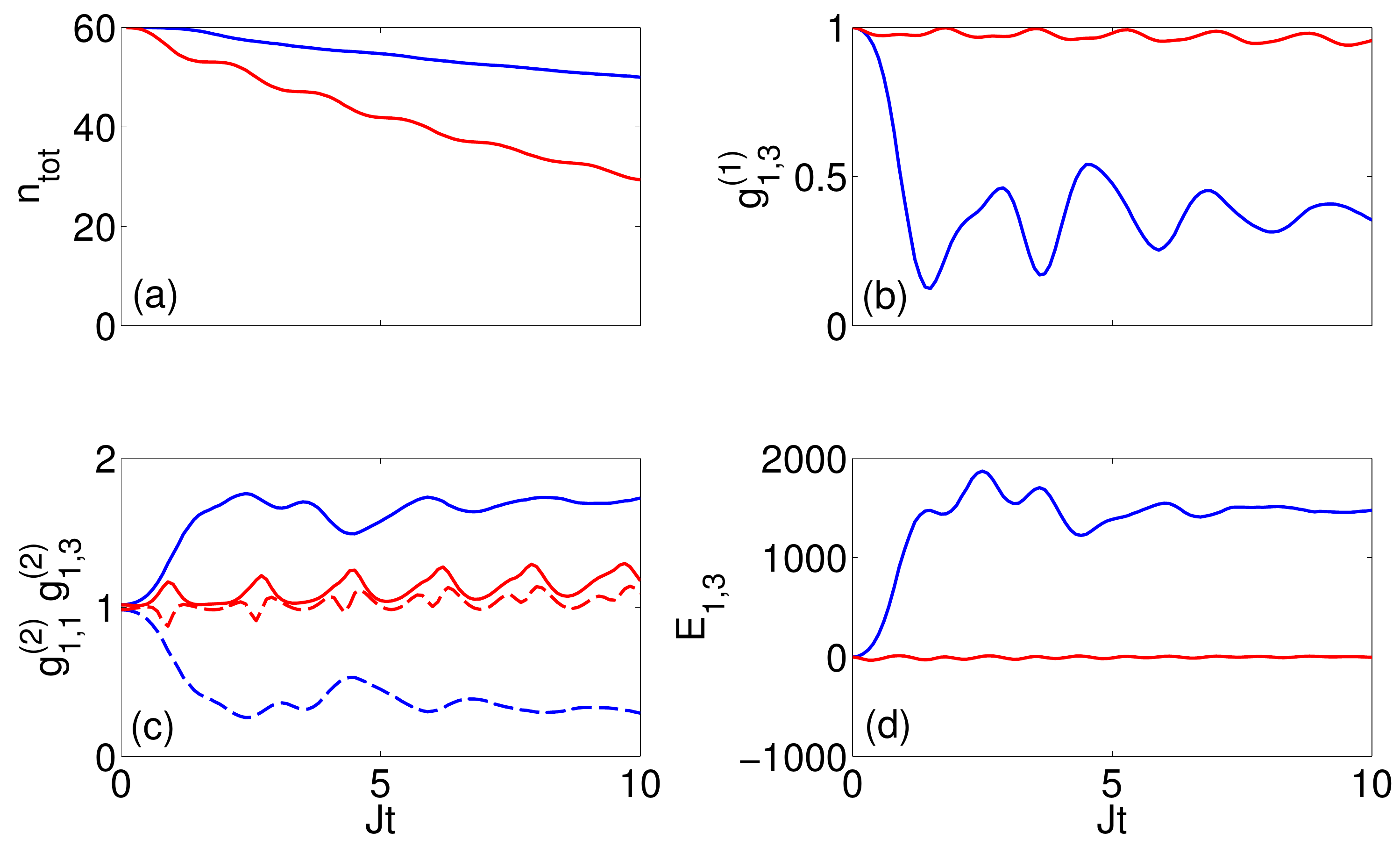}%
  \caption{\label{fig:03}
    Dynamics of the atom number, the correlation functions and the entanglement parameter in an
open Bose-Hubbard trimer with loss from site 2 for strong interactions, $U=0.1J$.
Plotted are (a) the total particle number $n_{\rm tot}$, (b) the phase coherence between
the sites 1 and 3 $g^{(1)}_{1,3}$, (c) the number correlations between
the sites 1 and 3 $g^{(2)}_{1,3}$ (dashed lines) and the number fluctuations $g^{(2)}_{1,1}$ (solid lines), 
and (d) the entanglement parameter $E_{1,2}$. The dynamics are simulated for two initial states: A BEC
with symmetric wave function (red lines), a BEC with an anti-symmetric
wave function (blue lines). The loss rate is $\gamma_2 = 0.2 J$ and the initial populations are
$n_1(0)=n_3(0)=30$, $n_2(0)=0$ in both cases. Time is given in units of the single-particle tunneling time. 
For the simulations we used an exact {\em quantum jump} method averaging over 200 trajectories. Data
adapted from ref. \cite{kordas13}.}
\end{figure}

In Fig. \ref{fig:03} we simulated the dynamics for  strong onsite interactions ($UN \gg J$) in a Bose-Hubbard trimer with periodic boundary conditions and loss only from site 2. 
The trimer case was largely discussed in our previous publications \cite{kordas12, kordas13}. We briefly review the essence of these results since this case is paradigmatic but still relatively simple to grasp well the physical mechanism of dynamical stabilization by dissipation. We study the dynamics of two different initial states, a pure Bose-Einstein condensate with an (anti-) symmetric wavefunction for the trimer setup:
\begin{equation}
   |\Psi_\pm\rangle = \frac{1}{2^N \, \sqrt{N!}}
       (\hat{\alpha}_1^\dagger \pm \hat{\alpha}_3^\dagger  )^N |0\rangle.
       \label{eq:states}
\end{equation}
For the simulations we used an exact unravelling of the master equation based on the {\em quantum jump} method
from which the density matrix and all relevant observables can be reconstructed \cite{mol93, kordas12, kordas13}.
In Fig. \ref{fig:03} (a), we plot the evolution of the total particle number $n_{\rm tot}$. 
The anti-symmetric initial state $|\Psi_-\rangle$ is a stationary (so-called dark) state of the master equation (\ref{eq:DBH}) for $U=0$,
such that decay is absent in the non-interacting limit (cf.~\cite{kepe12}). Hence the decay is indeed very slow for this initial state.
This behavior originates from the destructive interference of atoms tunneling from the sites 1 and 3 to the leaky site 2. 
For strong interactions tunneling is still allowed but weak. Localized soliton-like states, also referred to as breathers, form
dynamically at the non-dissipative sites.

The dissipative dynamics drives the atoms to very different quantum
states depending on the initial state and the interaction strength $U$.
To characterize these states we analyze the first and second order
coherence or correlation functions between different sites of the lattice.
These functions are defined as
\begin{equation}
\label{eq:coh1}
  g^{(1)}_{j,k} = \frac{|\langle  \hat{\alpha}_j^\dagger \hat{\alpha}_k \rangle| }{
                            \sqrt{ \langle \hat n_j \rangle \langle \hat n_k \rangle}},
\end{equation}
and
\begin{equation}
\label{eq:coh2}
  g^{(2)}_{j,k} = \frac{ \langle  \hat n_j \hat n_k \rangle }{
                             \langle \hat n_j \rangle \langle \hat n_k \rangle},
\end{equation}
respectively. The coherences between the wells $j=1$ and $k =3$ are plotted in Fig. \ref{fig:03}(b)
for both initial states. The symmetric initial state  $|\Psi_+\rangle$ is stable 
and the BEC remains approximately pure during the temporal evolution. 
Particle dissipation can even increase the purity and coherence of the many-body
state. This counter-intuitive effect is discussed for a series of different dissipation protocols 
in \cite{witt08,witt09,witt11,trim11}. On the other hand, the anti-symmetric state $|\Psi_-\rangle$ is 
stable only provided that interactions are weak. For $U=0.1 J$, a sharp decrease of the first-order
coherence is observed. This indicates the dynamical destruction of the condensate. 

For $j = k$, Eq. (\ref{eq:coh2}) reduces to the normalized fluctuations of the number operator in the $j$th well:
$\langle  \hat n_j ^2\rangle/\langle  \hat n_j \rangle^2$. The evolution of the second-order coherences are shown in Fig. \ref{fig:03} (c).
These quantities are essentially constant for a condensate with a symmetric initial wave function $|\Psi_+\rangle$.
On the other hand, for strong interactions, strong anti-correlations manifest for the initial state $|\Psi_-\rangle$.
This results in $g^{(2)}_{1,3} \ll 1$, implying a bunching of the atoms in exactly one of the non-dissipative wells, while the other two sites remain
essentially empty. It turns out that the two contributions localized either at site 1 or 3 remain \emph{coherent} \cite{kordas12}. 
As shown below, the atoms relax deterministically to a macroscopically entangled state of many atoms, 
reminiscent of the famous Schr\"odinger cat state (cf.~\cite{Leib05}).
Like in the previous section, we refer to such states as breather states as they correspond to the discrete breathers in extended
lattices in the semiclassical limit \cite{Camp04,Flac08}.

The particles in a breather state are strongly entangled. This implies that if some particles are 
measured at one site, the remaining atoms will be projected onto the same well with large probability.
To unambiguously detect the corresponding multi-particle entanglement, we use an entanglement
witness introduced in \cite{kordas12} (see the appendix therein). This entanglement parameter
generalizes the two-mode squeezing parameter from ref. \cite{zoller2001} and has the advantage of
being accessible in the experiment \cite{mueller2010}. Since we are using quantum jump simulations \cite{mol93}, we can
assume that our quantum state is decomposed into pure states, $\hat \rho = M^{-1} \sum_{\ell=1}^M |\psi_\ell\rangle\langle\psi_\ell |$.
The mode-entanglement parameter introduced in \cite{kordas12} is then
defined as
\begin{eqnarray}
         \label{eqn:ent_para}
\nonumber    E_{1,3} &:=& \langle (\hat{n}_1 - \hat{n}_3)^2 \rangle
     - \langle \hat{n}_1 - \hat{n}_3 \rangle^2
     - \langle \hat{n}_1 + \hat{n}_3 \rangle  \\
    &&  - \frac{1}{2M^2} \sum_{\ell,j}
     \left[ \langle (\hat{n}_1 - \hat n_3) \rangle_\ell
         - \langle (\hat{n}_1 - \hat{n}_3) \rangle_j \right]^2,
\end{eqnarray}
where $\langle \cdot \rangle_{\ell, j}$ denotes the expectation value in the pure state $|\psi_{\ell, j}\rangle$.  
The final term in the parameter $E_{1,3}$ corrects for the possibility of an incoherent superposition of states localized at sites $1$ and $3$.
As done in the appendix of ref. \cite{kordas12}, for a separable quantum state one can show that
$E_{1,3} \leq 0$, such that a value $E_{1,3}>0$ proves entanglement of the particles. 

Fig. \ref{fig:03} (d) shows the evolution of $E_{1,3}(t)$ for our two initial states. While the symmetric state $|\Psi_+\rangle$ remains
close to a pure BEC, such that $E_{1,3}(t) \approx 0$ for all times, the anti-symmetric state $|\Psi_-\rangle$ relaxes to the predicted strongly entangled breather state for strong interactions. In the latter case, the entanglement parameter grows to large values and saturates around $E_{1,3}(t) \approx 1500$. The breather states formed are metastable such that the generated entanglement does not persist forever but for reasonably long times relevant in the experiment. In consequence, we predict the formation of very stable many-body states characterized by large entanglement due to localized particle loss in the regime of strong interactions.

In the following, we investigate the breather formation in an extended optical lattice for two large systems with $M=40$ and $M=60$ sites. 
This will show that the formation of stable many-body states is not sensitive to the system size, which is interesting for their experimental realization since typically ten to a hundred lattice sites are effectively populated by the condensate. We apply periodic boundary conditions and dissipation occurs only at the lattice site $j=1$ for simplicity. As an initial state we use a pure BEC which is moved at constant speed \cite{Sias07} or accelerated \cite{Peik97} to the edge of the first Brillouin zone. Such a state with maximal phase change between neighboring sites corresponds to the antisymmetric state from Eq. (\ref{eq:states}). We consider the case of large filling factors, with $N/M=1000$. For such large systems, we can only use the truncated Wigner approximation for our numerical simulation. The method is reviewed in appendix B of ref. \cite{kordas13}. It includes beyond mean-field correlations to some extent in order to characterize the system's coherence.

\begin{figure}
  \includegraphics[width=\columnwidth]{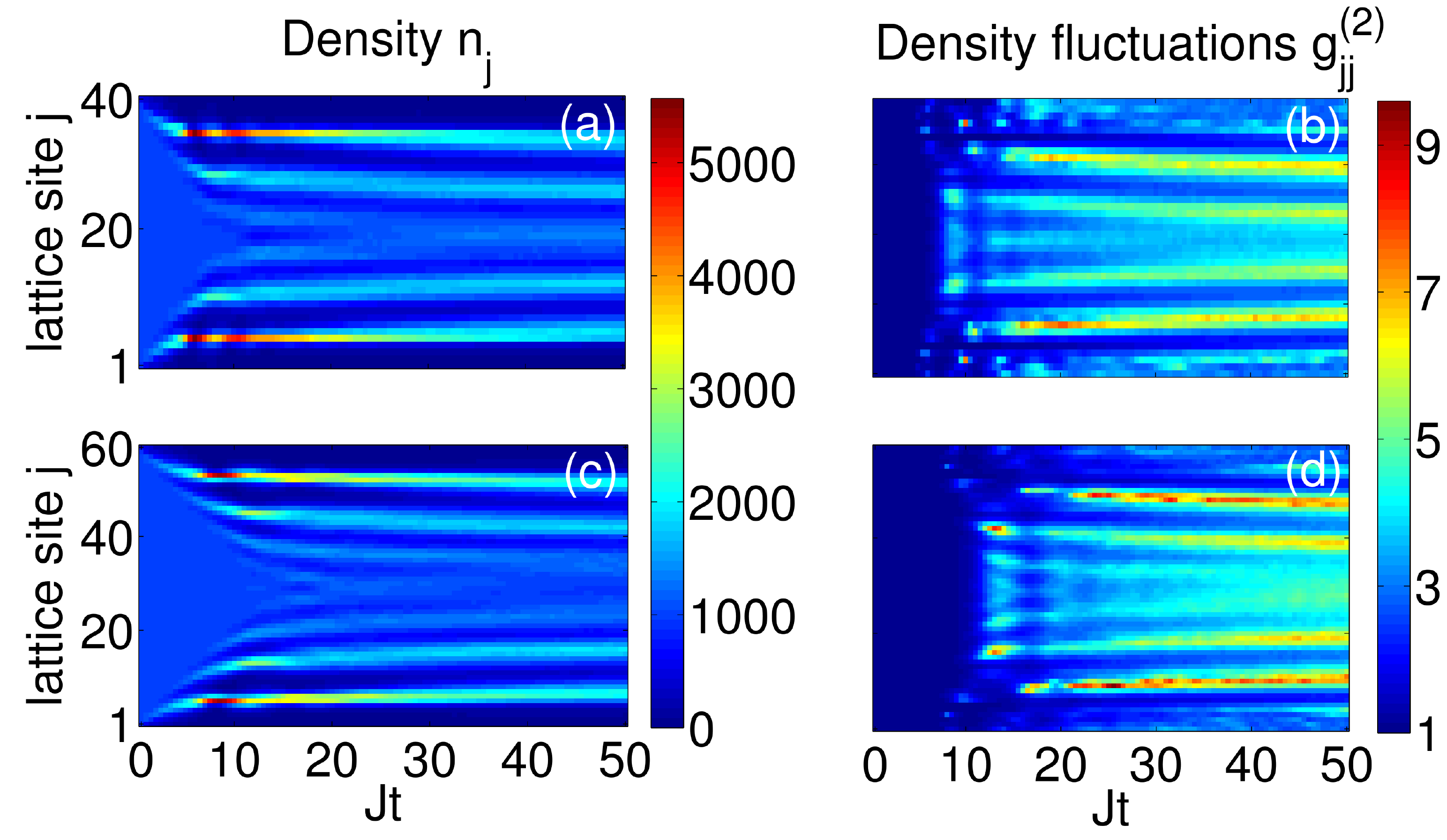}%
  \caption{\label{fig:04}
    Dynamics of a leaky Bose-Hubbard chain with $M=40$ (a,b) and 60 (c,d) wells. 
    We plotted (a,c) the particle density $\langle \hat n_j(t) \rangle$,
(b,d) the density fluctuations $g^{(2)}_{j,j}(t)$ in each lattice site.
For the simulations we used the {\it truncated Wigner} approximation
and the parameters $Un_{{\rm tot}}(0)=25 J$, $\gamma_1=2 J $, and $\rho(t=0)=n_{{\rm tot}}(0)/M=1000$.}
\end{figure}

\begin{figure}
  \includegraphics[width=\columnwidth]{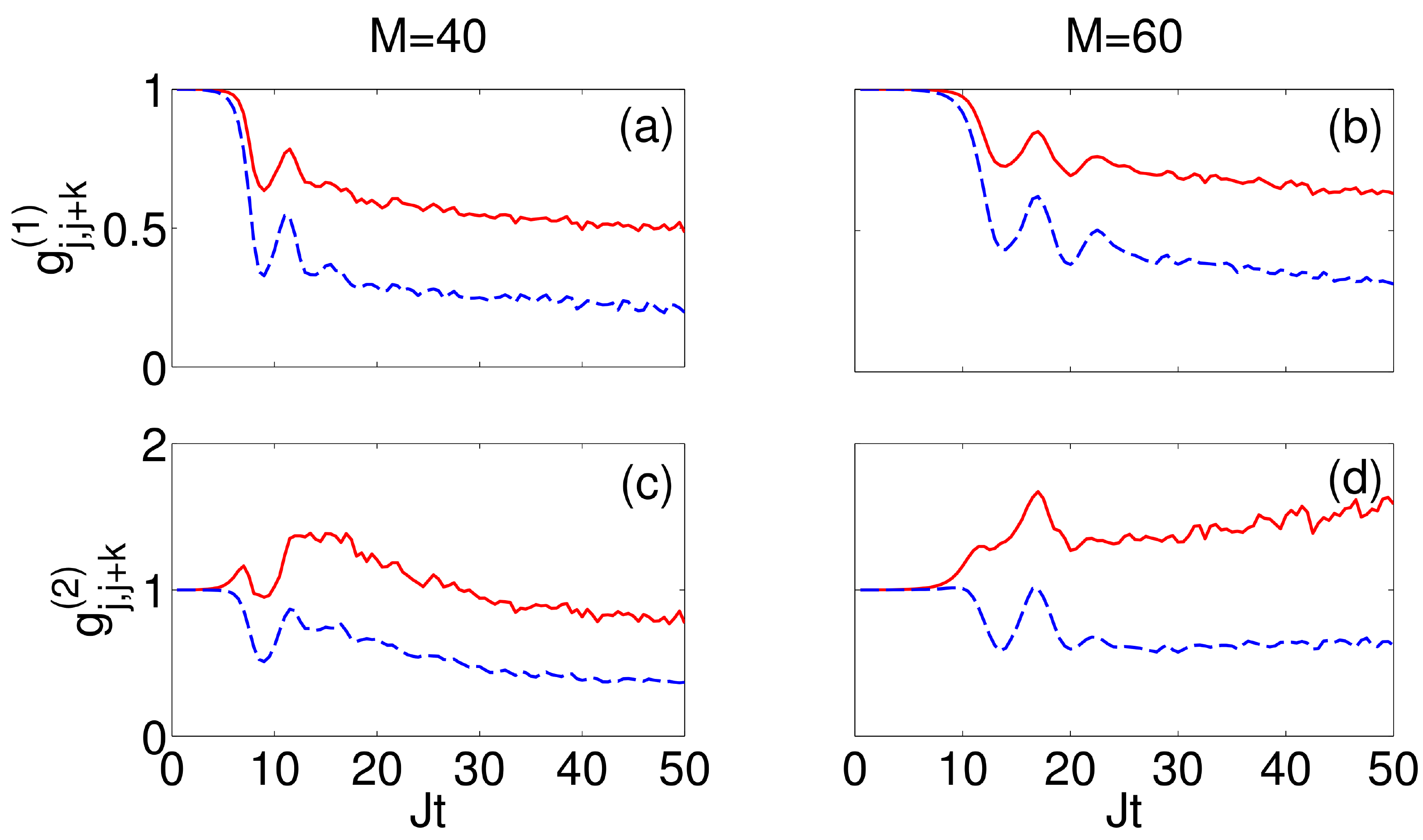}%
  \caption{\label{fig:05}
    For the dynamics shown in the previous figure, we present
    (a,b) the phase coherence $g^{(1)}_{j,j+k}(t)$
   and (c,d) the density-density correlations $g^{(2)}_{j,j+k}(t)$
   between site $j=20$ (a,c) or $j=30$ (b,d) and the corresponding neighboring sites with
   $k=1$ (solid red line) and $k=2$ (dashed blue line).}
\end{figure}

For weak interactions, the time-evolved state remains close to a pure BEC during the dissipation, such that all coherence functions remain approximately one. The evolution changes for strong interactions as shown in Figs.~\ref{fig:04} and ~\ref{fig:05}. Here, the phase coherence $g_{j,k}^{(1)}$ between neighboring sites is quickly lost, indicating a dynamical instability of the condensate. The second-order coherences $g_{j,j}^{(2)}$, however, rapidly increase (see Fig.~\ref{fig:04} (b)). This increase indicates a strong spatial bunching of the atoms in the same way as seen for the trimer above. Please note that the observed bunching in the dissipative equilibrium state is in strong contrast to the non-dissipative case, where the repulsive interactions tend to suppress number fluctuations in thermal equilibrium. Also strong anti-correlations with $g_{j,j+2}^{(2)} \approx 0.5$ are observed between the sites $j=20$ and $j=30$, respectively, and the corresponding next-to-nearest neighbor ones (see Fig.~\ref{fig:05}).

No such anti-correlations are found for the direct neighbor, as breathers typically extend over more than one site in our extended lattice geometry. In the same way as discussed above for the trimer case, the atoms tend to bunch at one well of the lattice, leaving the neighboring sites essentially empty. The exact position where the individual breathers form depends on the system parameters and their formation history during the dissipative time evolution.
The global many-body state is a coherent superposition of breathers at different lattice sites, just as the cat-like states formed in the trimer.
We conclude that the qualitative behavior of the dynamical breather formation is independent of the system size. Based on their versatile electron gun \cite{Ott08, Ott09, Ott2013, Ott_new}, the Kaiserslautern group could readily observe our predictions for breather formations in a quasi-one-dimensional lattice geometry and using as initial state either a band-edge or a phase-randomized (Mott) state.

\section{Refilling dynamics in the presence of noise and interactions}\label{sec:05}

Inspired by ongoing experiments on the non-equilibrium dynamics in periodic lattice structures \cite{Ott_new},
we are now looking at the problem of refilling an initially empty lattice site by atoms tunneling from the neighboring sites.
Our model is sketched in Fig.~\ref{fig:01} (b). We restrict to a minimal model of three potential wells in order to exactly unravel the master equation (\ref{eq:DBH}), with global uniform phase noise, see Eq.~(\ref{eq:L-noise}) but no dissipation. Without dissipation the mean-field approaches used above would be less reliable, therefore we compute the dynamics using the quantum jump method as used in section \ref{sec:04}, see Fig.~\ref{fig:03}. In an experiment, the central site is emptied with an electron beam \cite{Ott08, Ott09, Ott2013}, while initially the lattice is so deep such as to freeze tunneling completely \cite{Ott_new}. This motivates our use of random phase initial states with 15 bosons in the left and right well, respectively. For the evolution, the lattice can then be ramped down in the experiment in order to control the many-body tunneling rates \cite{Ott_new}.

\begin{figure}
  \includegraphics[width=\columnwidth]{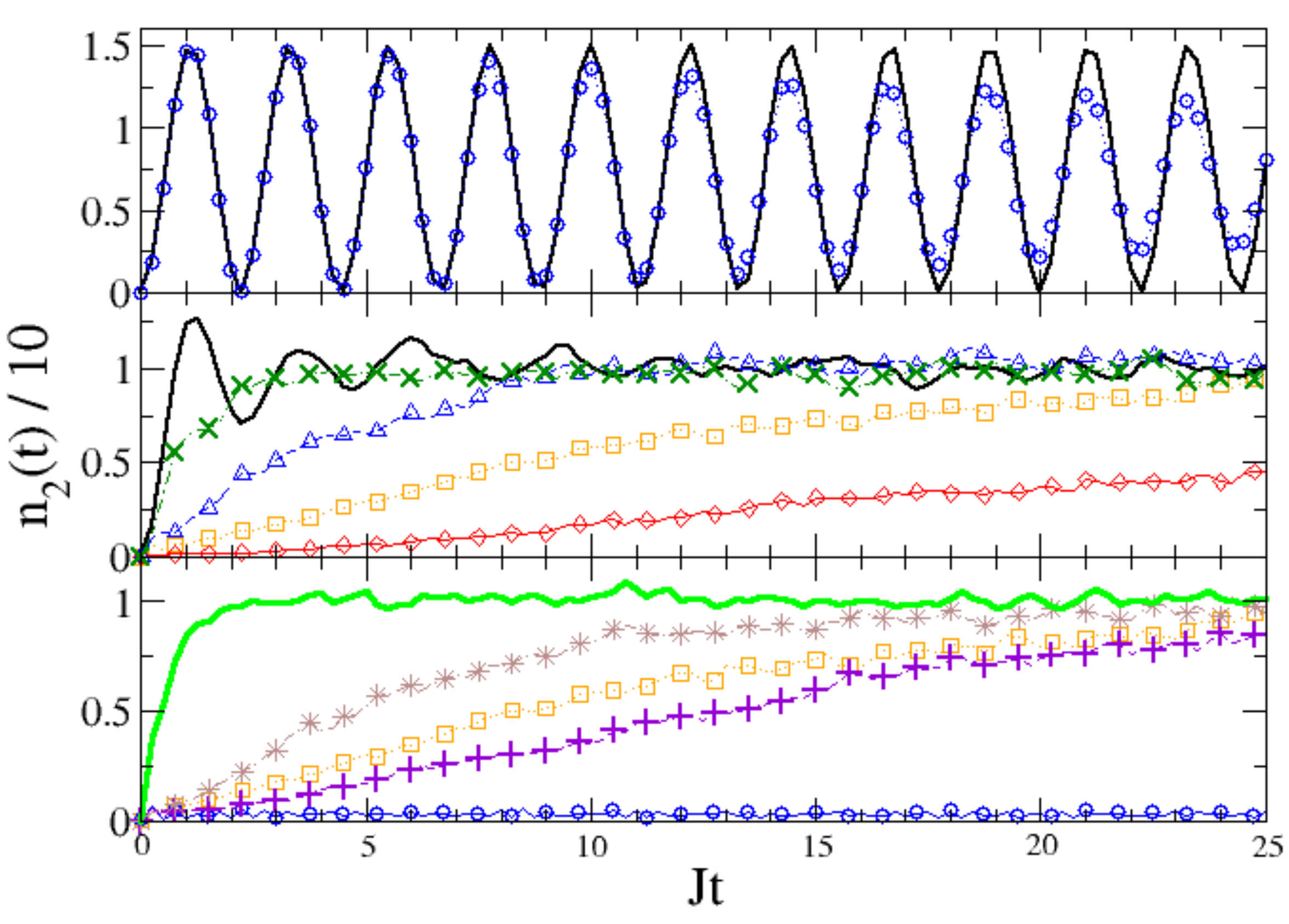}%
  \caption{\label{fig:06}
    Refilling dynamics of the central well shown in Fig. \ref{fig:01}(b) for a variety of parameters at fixed $J=1$. 
    The total number of atoms is 
    $N=30$, with 15 each in the two outer wells initially. Upper panel: $\kappa=0$ and $U=0$ (solid black line) or $UN=0.5$ (blue circles). 
    Middle panel:
    $\kappa=1$ and $U=0$ (solid black line), $UN=10$ (green crosses), $UN=20$ (blue triangles), $UN=30$ 
    (orange squares), $UN=50$ (red diamonds). Lower panel: $UN=30$ and $\kappa=0$ (blue circles), $\kappa=0.5$ (violet plusses), 
    $\kappa=1$ (orange squares), $\kappa=3$ (brown stars), and $\kappa=6$ (green thick line), respectively.
    }
\end{figure}

Our results are shown in Fig.~\ref{fig:06} for a fixed total number of atoms $N=30$ and $J=1$. The filling dynamics
strongly depends on the interactions $UN$ and on the strength of the phase noise $\kappa$.  Initially, the system is out of equilibrium. The dynamics enforces the approach of a new equilibrium state corresponding for $t \to \infty$ to an equal population distribution, corresponding to a filling of 10 in our case. The coherent oscillations visible in the populations, e.g. in the noise-free case (upper panel of Fig.~\ref{fig:06}) are strongly damped by the dephasing. In all cases eventually the populations will be equally distributed between the three sites on average, yet the filling times depend strongly on the interactions and the phase noise which act against each other. 
The former tend to produce self-trapping, i.e. they create a large difference in the chemical potentials between the empty site and the filled ones. The phase noise indeed helps the atoms to jump randomly into the central well step by step, thus enhancing the tunneling.
This simple but experimentally, in principle, easily accessible model represents a first step toward true particle transport along a chain of potential wells as studied now in the next section.


\section{Non-equilibrium neutral particle transport}\label{sec:06}

We now model two atom reservoirs by using an appropriate creation and destruction of particles in the outer sites of a lattice. As in refs. \cite{Pizo13, Ivan13} we are interested explicitly in the transport of neutral atoms, not in heat conductance across the sample as studied, e.g. in refs. \cite{anglin2011}. Our system consists of a lattice, with $M+2$ sites, where the outer two wells have a sink and a source of single particles with rates $\Gamma_j(1+N_j)$ and $\Gamma_j N_j$, respectively (see Fig. \ref{fig:01}(c)). These simultaneous destruction and creation of particles with these specific rates results in an approximately constant particle number in the outer wells during the whole evolution.

Indeed it is easy to see that the population of the outer wells remain constant.
We begin with the exact evolution equations for the population of the outer two wells:
\begin{eqnarray}
\frac{d}{dt}\langle \hat{\alpha}_L^\dagger \hat{\alpha}_L \rangle &=& -2J \Im\langle\hat{\alpha}_L^\dagger \hat{\alpha}_1\rangle
 - \frac{\Gamma_L}{2}\langle\hat{\alpha}_L^\dagger \hat{\alpha}_L\rangle + \frac{\Gamma_L}{2}N_L,\\
\frac{d}{dt}\langle \hat{\alpha}_R^\dagger \hat{\alpha}_R \rangle &=& -2J \Im\langle\hat{\alpha}_R^\dagger \hat{\alpha}_M\rangle
- \frac{\Gamma_R}{2}\langle\hat{\alpha}_R^\dagger \hat{\alpha}_R\rangle + \frac{\Gamma_R}{2}N_R,
\end{eqnarray}
if we assume $J\ll \Gamma_{L,R}$ we can approximate the above equations as follows
\begin{eqnarray}
\frac{d}{dt}\langle \hat{\alpha}_L^\dagger \hat{\alpha}_L \rangle &\approx& - \frac{\Gamma_L}{2}\langle\hat{\alpha}_L^\dagger \hat{\alpha}_L\rangle + \frac{\Gamma_L}{2}N_L,\\
\frac{d}{dt}\langle \hat{\alpha}_R^\dagger \hat{\alpha}_R \rangle &\approx& - \frac{\Gamma_R}{2}\langle\hat{\alpha}_R^\dagger \hat{\alpha}_R\rangle + \frac{\Gamma_R}{2}N_R.
\end{eqnarray}
The above equations have the following analytical solutions
\begin{eqnarray}
\langle \hat{\alpha}_L^\dagger \hat{\alpha}_L \rangle_{t} &\approx& (\langle\hat{\alpha}_L^\dagger \hat{\alpha}_L \rangle_{0}-N_L)e^{-\frac{\Gamma_L}{2}t} + N_L,\\
\langle \hat{\alpha}_R^\dagger \hat{\alpha}_R \rangle_{t} &\approx& (\langle\hat{\alpha}_R^\dagger \hat{\alpha}_R \rangle_{0}-N_R)e^{-\frac{\Gamma_R}{2}t} + N_R,
\end{eqnarray}
now if we initially have $\langle\hat{\alpha}_L^\dagger \hat{\alpha}_L \rangle_{0}=N_L$ and $\langle\hat{\alpha}_R^\dagger \hat{\alpha}_R \rangle_{0}=N_R$
then in the whole evolution of our system the particle number will remain almost constant
\begin{equation}
 \langle\hat{\alpha}_L^\dagger \hat{\alpha}_L \rangle_{t}\approx N_L,~\langle\hat{\alpha}_R^\dagger \hat{\alpha}_R \rangle_{t}\approx N_R.
\end{equation}
This very simple model allow us to control the population of the outer two wells in order to create a ``voltage''
between the two ends of the lattice.

Now that we have set up our model, we would like to know how the transport properties change when we vary the parameters, in particular, the interparticle interactions. For our numerical calculations, we use the truncated Wigner method, c.f. Fig. \ref{fig:04}.  As we are going to see, when a steady state is reached, the system loses its coherence, which can be characterized well by this beyond mean-field method, see Fig. \ref{fig:08} below. We are interested in particle transport, so it is natural to introduce the following particle current operators
\begin{eqnarray}
 \hat{j}_L &=& iJ(\hat{\alpha}_1^\dagger \hat{\alpha}_L - \hat{\alpha}_L^\dagger \hat{\alpha}_1),\\
\hat{j}_R &=& iJ(\hat{\alpha}_R^\dagger \hat{\alpha}_M - \hat{\alpha}_M^\dagger \hat{\alpha}_R).
\label{cur}
\end{eqnarray}
Thus the current from the left reservoir to the chain and the current from the chain to the right reservoir are given by the expressions
\begin{eqnarray}
 j_L &\equiv& \langle\hat{j}_L\rangle = -2J\Im\langle \hat{\alpha}_1^\dagger \hat{\alpha}_L\rangle,\\
 j_R &\equiv& \langle\hat{j}_R\rangle = -2J\Im\langle \hat{\alpha}_R^\dagger \hat{\alpha}_M\rangle.
\label{flux1}
\end{eqnarray}
We have defined the currents in such a way that they will be both positive if the particles flow from the left reservoir to the right one.
Then the following continuity equation holds
\begin{equation}
 \hat{j}_L - \hat{j}_R=\partial_t \hat{n}_{tot} \equiv \partial_t (\hat{n}_1+...+\hat{n}_M).
\label{contin}
\end{equation}
\begin{figure}
  \includegraphics[width=\columnwidth]{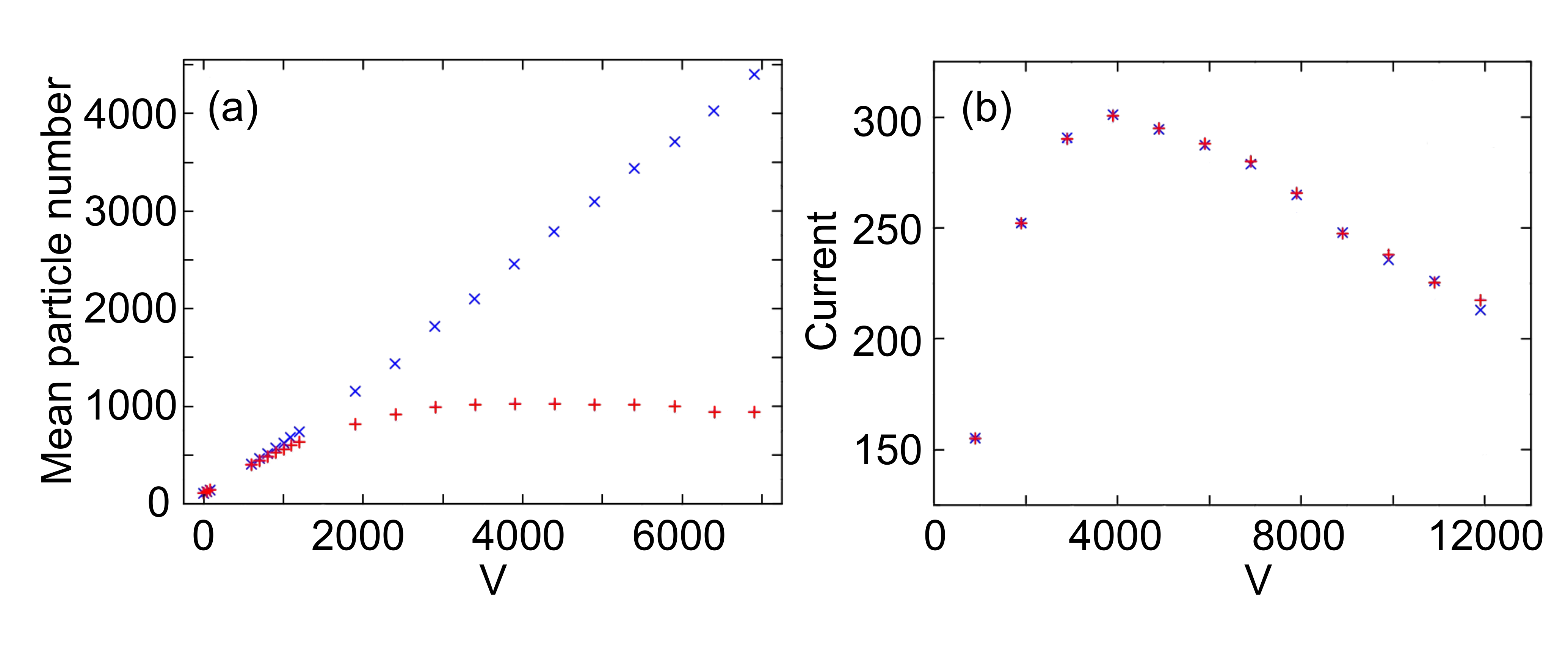}%
  \caption{\label{fig:07}
    (a) The mean particle number in each lattice site (blue x for the first well, red crosses for the second)
and (b) the current (blue x for $j_L$ and red crosses for $j_R$) as a function of the voltage $V=N_L-N_R$,
after the steady state has been reached.
The parameters are $U=10^{-3}J$ and $\Gamma_L=\Gamma_R=10J$. For the simulations we used the {\em truncated Wigner} approximation.}
\end{figure}

Let us  observe what happens when we change the ``voltage'', given the particle number difference of the reservoirs $V= N_L-N_R$.
Our system consists of four wells from which the outer two wells model the reservoirs or leads. Experimentally, one may think of Bose condensates with very large particle numbers, which are coupled to the inner geometrically constrained sites of the lattice. Fig. \ref{fig:07} (a) shows the particle numbers in the two wells as a function of voltage in the interacting case. We observe two regimes: in the first one, for $V\lesssim 1000$, both particle numbers are the same, and they increase linearly with the voltage. In the second regime, for $V\gtrsim 1000$, the particle number in the first well (the well that is connected with the reservoir with the larger particle number) increases linearly with the voltage, while in the second well the particle number becomes almost constant. This is again a consequence of the self-trapping effect: the change of behavior appears when the macroscopic interaction strength is greater than the tunneling strength, $Un_{\rm tot}(0)>J$.

Now, let us discuss the behavior of the steady state current. In Fig. \ref{fig:07}(b) we observe that the current, as a function of the voltage, has a maximum after which it drops again. This means that for large
voltages the transport of the particles through the lattice is blocked.
The same qualitative behavior was observed in~\cite{Ivan13} by approximating the interaction contribution to the self-energy
by the tabpole diagram, in the non-equilibrium Green's function framework.
The blockade is a consequence of the interactions:
as we saw in Fig. \ref{fig:07}(a) the particle number in the first lattice site increases with the voltage,
which also means that the macroscopic interaction strength increases in that lattice site.
Thus, we can conclude that the strong interactions that appear in the first lattice site block the transport of the particles.
Finally, we plot the coherence functions introduced in Eqs. (\ref{eq:coh1}) and (\ref{eq:coh2}). They show that the transporting current across the lattice system is essentially incoherent, see Fig. \ref{fig:08}. Hence, the incoherent reservoirs pump their coherence properties into the lattice site, resulting in an essentially incoherent steady-state transport scenario. Only if the
interactions are strong enough, just like in section \ref{sec:04}, metastable breather states tend to form inside the lattice, stabilized energetically by self-trapping.

\begin{figure}
  \includegraphics[width=\columnwidth]{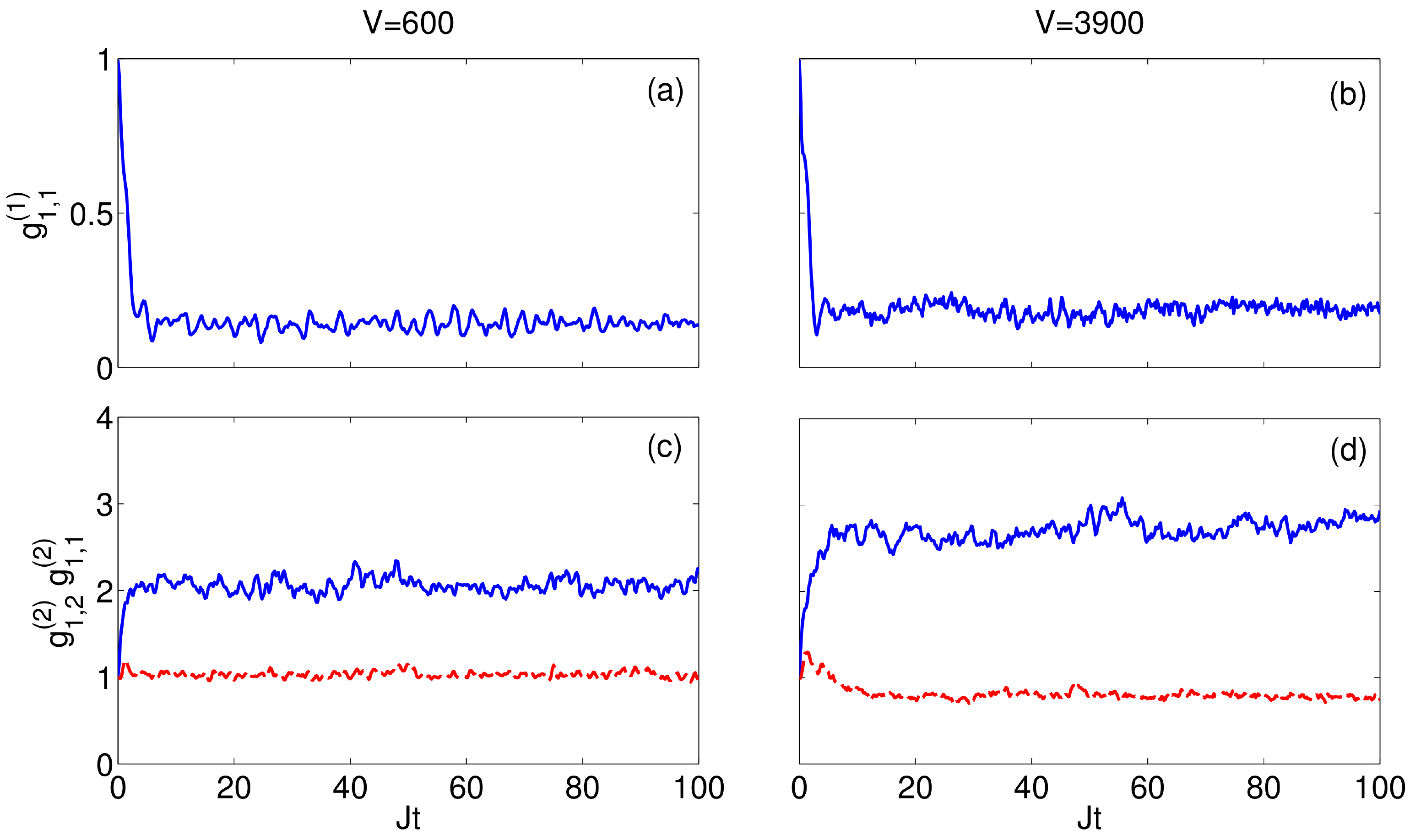}%
  \caption{\label{fig:08}
    (a,b) first and (c,d) second order coherence functions for two different voltages $V$ as indicated above the panels.
    Parameters and simulation method are the same as in the previous figure.  
    Left panels: the system relaxes in a thermal-like state, with $g^{(1)}_{1,1}$ close to zero, while $g^{(2)}_{1,2} \sim 1$ and  $g^{(2)}_{1,1}\sim 2$. Right panels: 
    For stronger voltage differences, and hence stronger self-trapping, we observe that $g^{(2)}_{1,1}\sim 3$ and weak  
    antibunching remains with $g^{(2)}_{1,2}<1$. 
    }
\end{figure}

\begin{figure}
  \includegraphics[width=0.9\columnwidth]{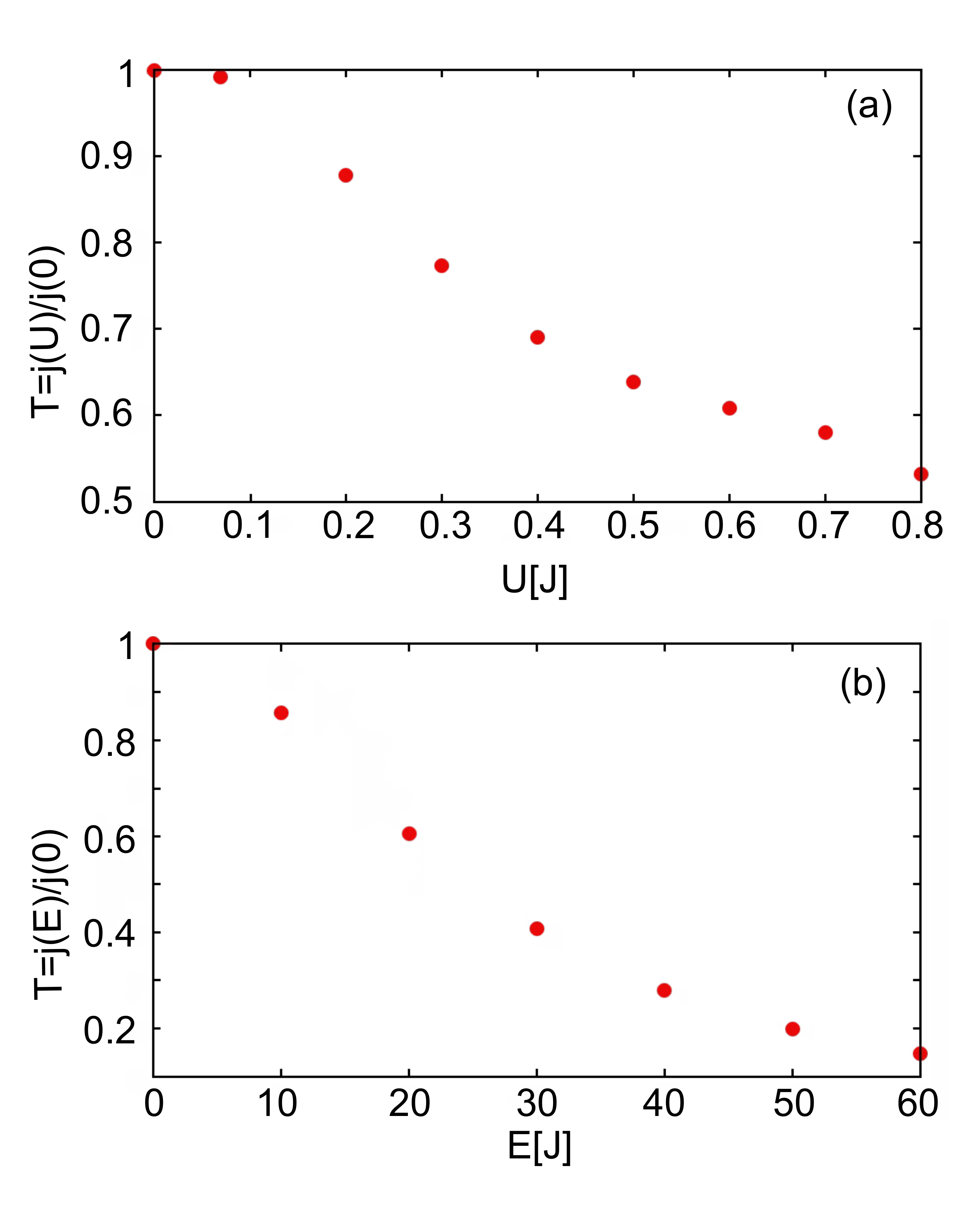}%
  \caption{\label{fig:09}
(a) The transmission coefficient through the central site, in which interactions appear, as a function of the interactions on this site.
The lattice is flat with initial populations $n_L(0)=15$ and $n_R(0)=5$.
(b) The transmission coefficient through the non-resonant site as a function of the detuning in the non-interacting case.
   $U=0$ everywhere, with initial populations $n_L(0)=30$ and $n_R(0)=10$.
   The other parameters are $\Gamma_L=\Gamma_R=50J$ and $n_1(0)=n_2(0)=n_3(0)=10$.}
\end{figure}

In order to understand better the effect of interactions and self-trapping, we now study the transport through a Bose-Hubbard chain but with interactions only in one lattice site. The system consists of five wells: the outer two are the reservoirs while the interactions are everywhere zero except for the central site where interactions are present. In Fig. \ref{fig:09}(a) we have the transmission coefficient, that is the steady state current through the middle interacting site divided by the current through the middle site when the interactions would be absent, $U=0$, as a function of the interaction strength $U$. As one can see the transmission coefficient drops as the interactions are increased. This behavior
confirms the observation from above. The interactions in the central site act as a barrier,
which blocks the transport of the atoms through the middle interacting site.
The particles from the reservoir are forced to enter the interacting site but there they are trapped, since
they cannot get rid of the energy by tunneling to the neighboring sites. As the particle number
increases it becomes harder and harder for the particles  to tunnel out of the interacting site, since only this site
is out of resonance. 

To model the inhibition of transport by the interaction-induced energy gap, we now consider an effective model, for which
we detune from resonance the energy level of the central site, with zero interactions everywhere. Since there are no interactions in our system we can write down the exact evolution equations for the elements of the single particle density matrix $\sigma_{j,k}\equiv\langle \hat{\alpha}_j^\dag \hat{\alpha}_k\rangle$. These equations can be obtained from the master equation via
$\dot{\sigma}_{j,k}={\rm Tr}[\hat{\alpha}_j^\dag \hat{\alpha}_k\dot{\hat{\rho}}]$, see also \cite{witt11}, resulting in
\begin{eqnarray}
\nonumber i\dot{\sigma}_{j,k} &=& (\varepsilon_k-\varepsilon_j)\sigma_{j,k}\\
 \nonumber && - J \left(\sigma_{j,k+1}
+ \sigma_{j,k-1} - \sigma_{j+1,k} - \sigma_{j-1,k}\right) \\
\nonumber && -\frac{i}{2}\left( \Gamma_j(\delta_{j,L}+\delta_{j,R}) + \Gamma_k(\delta_{k,L} + \delta_{k,R})\right)\sigma_{j,k}\\
&& + \frac{i}{2}\delta_{j,k}(\delta_{j,L}+\delta_{j,R})\Gamma_j N_j,
\end{eqnarray}
where $k,j=L,1,2,3,R$.

In Fig. \ref{fig:09}(b) we plot the transmission coefficient $T$, that is the steady state current through the middle non-resonant site divided by the current through the middle site of a flat lattice ($\varepsilon_2\equiv E=0$), as a function of the energy of the non-resonant site. The transmission coefficient decreases as the energy difference between the central well and the neighboring sites increases. This behavior confirms the argument we gave previously. A large energy difference between neighboring sites -- be it due to interactions or non-resonant detuning -- blocks the transport of the bosons. Both effects are well-known in electronic transport theory \cite{Qdots}, and a qualitatively similar behavior was observed in the mean-field limit for Bose condensates in tilted optical lattices \cite{Sias07,WS2005,PMW2013}.
The transport results presented here must be seen as a first step to propose experiments which realize neutral particle transport across lattice structures to allow for atomtronic applications as envisaged, e.g. in refs. \cite{Eckel2014, Cali2012, Holland2010}.


\section{Conclusions and perspectives}\label{sec:07}

We have shown that dissipation, dephasing and interparticle interactions can cooperate to produce coherent structures in optical lattices for ultracold bosons. Using appropriate initial states, we can deterministically produce also quantum superpositions of such highly populated states. The coupling to incoherent reservoirs leads to an essentially incoherent particle current across a chain of quantum dots. This current is suppressed by dynamical self-trapping due to interactions producing again very stable solitonic states of atoms. New questions for future research include the presence of disorder \cite{BW2008, Richter2014} and its impact on many-body quantum transport. Also the simultaneous presence of transport and coherent driving may be investigated, see e.g. \cite{Pizo13, Tomadin2012}.

On the theoretical front new methods for the approximate or possibly exact evolution of interacting bosons and fermions are heavily needed going along with the advance of experimental many-body quantum simulators \cite{Qsim2012}. This is true, in particular, in higher spatial dimensions and in the presence of true quantum mechanical environments in order to further engineer the complex dynamics of open many-body systems. The applicability of standard methods of many-body physics originally developed for detecting phase transitions is highly questionable. For instance, the time-dependent Gutzwiller ansatz used, e.g. in ref. \cite{hofstetter2014}, does not well describe the incoherent transport possibly induced by dissipative couplings. Also the time-dependent version of the density matrix renormalization group algorithm, see e.g. \cite{schollwoeck2011, kollath2011}, is applicable safely only in one spatial dimension and not dynamically stable for strongly interacting bosons \cite{daley2009}. Another method which generalizes the so-called Mori projector \cite{Mori} allows for the calculation of local properties of a closed or open lattice problem. However, its basic assumption that the initial state factorizes with respect to the subsystems makes the method difficult to apply in lattices filled with condensates, where this assumption is generally not valid. An interesting direction is finally that of coherent-state path integrals in the continuum \cite{Athen, Kordas2015}, which -- combined with the Feynman-Vernon theory -- could be used as the basis for systematic approximations in open lattice systems.



\begin{thebibliography}{0}

\bibitem{Zoller2008} 
S. Diehl, A. Micheli, A. Kantian, B. Kraus, H. P. B\"uchler, and P. Zoller, Nat. Phys. {\bf 4}, 878 (2008).

\bibitem{Cirac2009} 
F. Verstraete, M. M. Wolf, and J. I. Cirac, Nat. Phys. {\bf 5}, 633 (2009).

\bibitem{Nature_new} 
Y. Lin, J. P. Gaebler, F. Reiter, T. R. Tan, R. Bowler, 
A. S. S\/orensen, D. Leibfried, and D. J. Wineland, Nature {\bf 504},  415 (2013).

\bibitem{kordas12}%
  \textsc{G.~Kordas},
  \textsc{S.~Wimberger}, and
  \textsc{D.~Witthaut}, 
  {\jr EPL} \textbf{100}, 30007 (2012).
  
\bibitem{Plenio2013} 
A. W. Chin, J. Prior, R. Rosenbach, F. Caycedo-Soler, S. F. Huelga, and M. B. Plenio,
Nat. Phys. {\bf 9}, 113 (2013).

\bibitem{Brantut2012}
J. P. Brantut, J. Meineke, D. Stadler, S. Krinner, and T. Esslinger,
Science {\bf 337}, 1069 (2012).

\bibitem{Eckel2014}
S. Eckel, J. G. Lee, F. Jendrzejewski, N. Murray, C. W. Clark, C. J. Lobb, W. D. Phillips, 
M. Edwards, and G. K. Campbell, Nature {\bf 506}, 200 (2014).

\bibitem{Krinner2014}
S. Krinner, D. Stadler, D. Husmann, J. P. Brantut, and T. Esslinger, arXiv:1404.6400.

\bibitem{Ott_new}%
  \textsc{R. Labovie},
  \textsc{B. Santra},
  \textsc{S. Heun}, 
  \textsc{S.~Wimberger}, and
   \textsc{H. Ott}, 
  {\jr Phys. Rev. Lett.}. \textbf{115}, 050601 (2015).

\bibitem{BDZ2008}  
I. Bloch,  J. Dalibard, and W.  Zwerger, Rev. Mod. Phys. {\bf 80}, 885 (2008).

\bibitem{Anglin1997}%
\textsc{J. R. Anglin}, 
{\jr Phys. Rev. Lett.} \textbf{79}, 6 (1997). 

\bibitem{Zoller2010}%
H. Pichler, A. J. Daley, and P. Zoller, \jr{Phys. Rev. A} \textbf{82}, 063605 (2010). 

  \bibitem{Breu07}%
  \textsc{H.-P. Breuer}, and
  \textsc{F. Petruccione},
  {\it The Theory of Open Quantum Systems} (Oxford University Press, New York, 2007).

  \bibitem{witt08}%
  \textsc{D.~Witthaut},
  \textsc{F.~Trimborn}, and
  \textsc{S.~Wimberger},
  {\jr Phys. Rev. Lett.} \textbf{101}, 200402 (2008).

  \bibitem{witt09}%
  \textsc{D.~Witthaut},
  \textsc{F.~Trimborn}, and
  \textsc{S.~Wimberger},
  {\jr Phys. Rev. A} \textbf{79}, 033621 (2009).

  \bibitem{witt11}%
  \textsc{D.~Witthaut},
  \textsc{F.~Trimborn},
  \textsc{H.~Hennig},
  \textsc{G.~Kordas},
  \textsc{T.~Geisel}, and
  \textsc{S.~Wimberger},
  {\jr Phys. Rev. A} \textbf{83}, 063608 (2011).

  \bibitem{trim11}%
  \textsc{F.~Trimborn},
  \textsc{D.~Witthaut},
  \textsc{H.~Hennig},
  \textsc{G.~Kordas},
  \textsc{T.~Geisel}, and
  \textsc{S.~Wimberger},
  {\jr Eur. Phys. J. D} \textbf{63}, 63 (2011).

 \bibitem{kepe12}%
  \textsc{K.\,V.~Kepesidis}, and
  \textsc{M.\,J.~Hartmann},
  {\jr Phys. Rev. A} \textbf{85}, 063620 (2012).

\bibitem{kollath2011}
P. Barmettler and C. Kollath, {Phys. Rev. A} {\bf 84}, 041606 (2011).

\bibitem{Kordas2015}
G. Kordas, D. Witthaut, P. Buonsante, A. Vezzani, R. Burioni, A. I. Karanikas, and S. Wimberger, {Eur. Phys. J ST} {\bf 224},127 (2015).

  \bibitem{kordas13}%
  {G.~Kordas},
  {S.~Wimberger}, and
  {D.~Witthaut}, {Phys. Rev. A} \textbf{87}, 043618 (2013).

\bibitem{Ott08}
T.~Gericke,  P.~W\"urtz, D.~Reitz, T.~Langen, and H.~Ott,
Nat. Phys. {\bf 4}, 949 (2008).

\bibitem{Ott09}
P.~W\"urtz, T.~Langen, T.~Gericke, A.~Koglbauer, and H.~Ott,
Phys. Rev. Lett. {\bf 103}, 080404 (2009).

\bibitem{Laser}
C. Weitenberg, M. Endres, J. F. Sherson, M. Cheneau, P. Schauß, T. Fukuhara, I. Bloch, and 
S. Kuhr, Nature {\bf 471}, 319 (2011).

\bibitem{Ott2013}
G. Barontini, R. Labouvie, F. Stubenrauch, A. Vogler, V. Guarrera, and H. Ott,
Phys. Rev. Lett. {\bf 110}, 035302 (2013).

\bibitem{Aspect2006}
W. Guerin, J.-F. Riou, J. P. Gaebler, V. Josse, P. Bouyer, and A. Aspect, 
Phys. Rev. Lett. {\bf 97}, 200402 (2006).

\bibitem{Nat_Phys}
N. P. Robins, C. Figl, M. Jeppesen, G. R. Dennis, and J. D. Close, Nat. Phys. {\bf 4},  731 (2008).

\bibitem{Gattobigio2011}
G. L. Gattobigio, A. Couvert, B. Georgeot, and D. Guery-Odelin, 
Phys. Rev. Lett. {\bf 107}, 254104 (2011).

\bibitem{Cali2012}
S. C. Caliga, C. J. E. Straatsma, A. A. Zozulya, D. Z. Anderson, arXiv:1208.3109 (2012).

\bibitem{Camp04}%
  \textsc{D.\,K.~Campbell},
  \textsc{S.~Flach}, and
  \textsc{Y.\,S.~Kivshar},
  {\jr Phys. Today} \textbf{467}, 57 (2004).

  \bibitem{Flac08}%
  \textsc{S.~Flach}, and
  \textsc{A.\,V.~Gorbach},
  {\jr Phys. Rep.} \textbf{467}, 1 (2008).

  \bibitem{Leib05}%
  \textsc{D.~Leibfried},
  \textsc{E.~Knill},
  \textsc{S.~Seidelin},
  \textsc{J.~Britton},
  \textsc{R.\,B.~Blakestad},
  \textsc{D.\,B.~Hume},
  \textsc{W.\,M.~Itano},
  \textsc{J.\,D.~Jost},
  \textsc{C.~Langer},
  \textsc{R.~Ozeri},
  \textsc{R.~Reichle}, and
  \textsc{D.\,J.~Wineland},
  {\jr Nature} \textbf{438}, 639 (2005).


  \bibitem{mol93}%
  \textsc{K.~M\o{}lmer},
  \textsc{Y.~Castin}, and
  \textsc{J.~Dalibard},
  {\jr J. Opt. Soc. Am. B} \textbf{10}, 524 (1993).

\bibitem{zoller2001}
A. S. Sørensen, L.-M. Duan, J. I. Cirac, and P. Zoller, Nature {\bf 409}, 63 (2001).

\bibitem{mueller2010}
J. Esteve, C. Gross, A. Weller, S. Giovanazzi, and M. K. Oberthaler
Nature {\bf 455}, 1216 (2008);
T. M\"uller, B. Zimmermann, J. Meineke, J.P. Brantut, T. Esslinger, and H. Moritz,
Phys. Rev. Lett. {\bf 105}, 040401 (2010).

  \bibitem{Sias07}%
  \textsc{C.~Sias},
  \textsc{A.~Zenesini},
  \textsc{H.~Lignier},
  \textsc{S.~Wimberger},
  \textsc{D.~Ciampini},
  \textsc{O.~Morsch}, and
  \textsc{E.~Arimondo},
  {\jr Phys. Rev. Lett.} \textbf{98}, 120403 (2007).

  \bibitem{Peik97}%
  \textsc{E.~Peik},
  \textsc{M.\,B.~Dahan},
  \textsc{I.~Bouchoule},
  \textsc{Y.~Castin}, and
  \textsc{C.~Salomon},
  {\jr Phys. Rev. A} \textbf{55}, 2989 (1997).

  \bibitem{Ivan13}%
  \textsc{A.~Ivanov},
  \textsc{G.~Kordas},
  \textsc{A.~Komnik}, and
  \textsc{S.~Wimberger},
  {\jr Eur. Phys. J. B} \textbf{86}, 345 (2013).

  \bibitem{Pizo13}%
  \textsc{I.~Pi\v{z}orn},
  {\jr Phys. Rev. A} \textbf{88}, 043635 (2013).
 
 \bibitem{anglin2011}
 L. Gilz and J. R. Anglin, Phys. Rev. Lett. {\bf 107}, 090601 (2011);
 J. P. Brantut, C. Grenier, J. Meineke, D. Stadler, S. Krinner, C. Kollath, T. Esslinger, and A. Georges,
Science {\bf 342}, 713 (2013).

  \bibitem{Qdots}%
  L.P. Kouwenhoven, C.M. Marcus, P.L. McEuen, S. Tarucha, R.M. Westervelt, and N.S. Wingreen,
  {\it Electron transport in quantum dots}, Kluwer Series, E345, Proceedings of the NATO Advanced Study Institute on Mesoscopic Electron Transport, 105-214 (1997).
  
  \bibitem{WS2005}
  S. Wimberger, R. Mannella, O. Morsch, E. Arimondo, A. R. Kolovsky, and A. Buchleitner, 
  Phys. Rev. A {\bf 72}, 063610 (2005); 
  G. Tayebirad, R. Mannella, and S. Wimberger, Appl. Phys. B {\bf 102}, 489 (2011).
  
  \bibitem{PMW2013}
C. A. Parra-Murillo, J. Madro\~nero, and S. Wimberger, Phys. Rev. A {\bf 88},  032119 (2013); 
C. A. Parra-Murillo, J. Madro\~nero, and S. Wimberger, Phys. Rev. A {\bf 89},  053610 (2014);
C. A. Parra-Murillo, J. Madro\~nero, and S. Wimberger, Comp. Phys. Comm. {\bf 186}, 19 (2015);
C. A. Parra-Murillo, J. Madro\~nero, and S. Wimberger, Annals of Physics {\bf 527},  656 (2015).

\bibitem{Holland2010}
R. A. Pepino, J. Cooper, D. Meiser, D. Z. Anderson, and M. J. Holland,
Phys. Rev. A {\bf 82}, 013640 (2010);
L. H. Kristinsdottir, O. Karlstr\"om, J. Bjerlin, J. C. Cremon, P. Schlagheck, A. Wacker, and S. M. Reimann,
Phys. Rev. Lett. {\bf 110}, 085303 (2013).

  \bibitem{BW2008}
P. Buonsante and S. Wimberger, Phys. Rev. A {\bf 77}, 041606(R) (2008).

 \bibitem{Richter2014}
T. Engl, J. Dujardin, A. Arg\"uelles, P. Schlagheck, K. Richter, and D. Urbina, Phys. Rev. Lett. {\bf 112}, 140403  (2014).

\bibitem{Tomadin2012}
M. J. Hartmann, Phys. Rev. Lett. {\bf 104}, 113601 (2010);
A. Tomadin, S. Diehl, M. D. Lukin, P. Rabl, and P. Zoller, Phys. Rev. A {\bf 86}, 033821 (2012); 
A. Le Boité, G. Orso, and C. Ciuti, Phys. Rev. Lett. {\bf 110}, 233601 (2013).

\bibitem{Qsim2012}
I. Bloch, J. Dalibard, S. Nascimb\`ene, Nat. Phys. {\bf 8}, 267 (2012).

\bibitem{hofstetter2014}
I. Vidanovic, D. Cocks, and W. Hofstetter, Phys. Rev. A {\bf 89}, 053614 (2014).

\bibitem{schollwoeck2011}
U. Schollw\"ock, Ann. Phys. (NJ) {\bf 326}, 96 (2011).

\bibitem{daley2009}
H. Venzl, A. J. Daley, F. Mintert, and A. Buchleitner, Phys. Rev. E {\bf 79}, 056223 (2009).

\bibitem{Mori}
H. Grabert, {\em Projection Operator Techniques in Nonequilibrium Statistical Mechanics} (Springer-Verlag, Heidelberg, 1982);
P. Degenfeld-Schonburg and M. J. Hartmann, Phys. Rev. B {\bf 89}, 245108 (2014).

\bibitem{Athen}
G. Kordas, S. I. Mistakidis, and A. I. Karanikas, Phys. Rev. A {\bf 90}, 032104 (2014).

\end{thebibliography}
\end{document}